\documentclass[12pt,showpacs,showkeys,amsmath,amssymb]{revtex4}
\usepackage{amsmath,amsfonts,amsthm,amscd,amssymb,latexsym}
\usepackage{bm}
\usepackage{dcolumn}
\usepackage{graphicx}
\usepackage{epstopdf}
\usepackage{color}
\usepackage{epsf}
\usepackage{epsfig}
\usepackage{graphicx, epic, eepic, color}

\def\pmb#1{\setbox0=\hbox{$#1$}%
  \kern-.025em\copy0\kern-\wd0
  \kern.05em\copy0\kern-\wd0
  \kern-.025em\raise.0433em\box0}

\def\beq{\begin{equation}}
\def\eeq{\end{equation}}

\begin{document}

\title{Jacobi Fields and Tidal Effects in Kerr Spacetime}

\author{Carmen \surname{Chicone}$^{1}$}
\email{chiconec@missouri.edu}
\author{Bahram \surname{Mashhoon}$^{2,3}$}
\email{mashhoonb@missouri.edu}

\affiliation{$^1$Department of Mathematics, University of Missouri, Columbia, Missouri 65211, USA\\
$^2$Department of Physics and Astronomy, University of Missouri, Columbia, Missouri 65211, USA\\
$^3$School of Astronomy, Institute for Research in Fundamental
Sciences (IPM), P. O. Box 19395-5531, Tehran, Iran\\
}

\date{\today}

\begin{abstract}
General relativistic tidal equations are formulated with respect to the rest frame of a central gravitational source described by the Kerr gravitational field. Specifically, observers that are spatially at rest in the exterior Kerr spacetime are considered in detail; in effect,  these fiducial observers define the rest frame of the Kerr source. The generalized Jacobi equation for the free motion of a test particle is worked out within the Fermi coordinate system established along the world line of a fiducial observer at rest on the symmetry axis of the exterior Kerr background.  The nature of the solutions of these equations are investigated in connection with the phenomena associated with astrophysical jets. 
\end{abstract}


\maketitle

\section{Introduction}

Quasars and active galactic nuclei often exhibit puzzling features involving relativistic outflows. These jets frequently come in pairs and originate from a central source that is believed to be a massive rotating black hole surrounded with an accretion disk. Similar phenomena have been observed in our galaxy and are associated with microquasars, which are binary x-ray sources. Jets are presumably emitted along the rotation axis of a Kerr black hole, a circumstance that appears to be contrary to the strong gravitational attraction of the black hole. Nevertheless, the double-jet feature is consistent with the symmetry of the Kerr gravitational field under reflection about its equatorial plane. Observational data further indicate that in extragalactic jets, jet acceleration away from the black hole results in Lorentz factors that are much greater than unity. 

The motion of particles in a jet  would be naturally affected by magnetohydrodynamic forces as well as relativistic gravitational effects of general relativity~\cite{Punsly}. The purpose of the present work is to explore further the influence of the gravitational tidal acceleration of the Kerr black hole on the dynamics of astrophysical jets within the framework of general relativity (GR); see~\cite{Bini:2017uax, Mashhoon:2020tha} and the references cited therein.  Unless specified otherwise, in this paper we use gravitational units such that $c = G = 1$ throughout; moreover, Greek indices run from 0 to 3, while Latin indices run from 1 to 3, and the signature of the spacetime metric is +2.

The Kerr metric, $-ds^2 = g_{\mu \nu}dx^\mu dx^\nu$, where $ds$ is the spacetime interval for timelike world lines, can be expressed  in Boyer-Lindquist coordinates $x^\mu = (\tilde{t},r,\theta,\phi)$ as~\cite{Chandra, BON}
\begin{align}\label{K1}
\nonumber -ds^2 = {}& -\left(1-\frac{2 M r}{\Sigma}\right)\,d\tilde{t}\,^2 - \frac{4M a r}{\Sigma} \sin^2\theta\,d\tilde{t}\,d\phi +\frac{\Sigma}{\Delta}\,dr^2+\Sigma\, d\theta^2 \\
&+\left(r^2+a^2 + \frac{2 M a^2 r}{\Sigma}\,\sin^2\theta \right)\,\sin^2\theta\, d\phi^2,
\end{align}
where $M > 0$ and $a > 0$ are the mass and specific angular momentum of the Kerr source, $\Sigma=r^2+a^2\cos^2\theta$ and $\Delta=r^2-2Mr+a^2$. Kerr spacetime is asymptotically flat; in fact,  by passing to the limit $r \to \infty$, metric~\eqref{K1} reduces to the Minkowski spacetime metric expressed in spherical polar coordinates. Furthermore, Kerr gravitational field is stationary and axisymmetric with Killing vector fields $\partial_{\tilde t}$ and $\partial_\phi$. The spacetime is a  Kerr black hole for $a \le M$; otherwise, the Kerr source is a naked singularity. For a Kerr black hole, reference observers can be spatially at rest from the asymptotic region ($r \to \infty$) all the way down to the exterior of the stationary limit surface, where $g_{00} = 0$ or $\Sigma = 2\,M r$.  On the symmetry axis, $\theta = 0$ and the exterior solution is valid for $r$ in the range $M+(M^2-a^2)^{1/2} < r<\infty$. 

Astronomers regularly observe energetic outflows and other events relative to the rest frames of the gravitationally collapsed systems that are usually modeled by  Kerr gravitational fields. To describe theoretically  such phenomena relative to the rest frame of the source, we imagine the existence of reference observers that are spatially at rest in the exterior Kerr spacetime. Such observers in effect define the rest frame of the Kerr source. A detailed description of this family of observers as well as their adapted nonrotating (i.e., Fermi-walker) orthonormal tetrad frames is contained in~\cite{Bini:2017uax, Mashhoon:2020tha}. In particular, for the present study our fiducial observer is at rest at radial position $r$ along the rotation axis of the exterior Kerr spacetime. The observer has unit 4-velocity vector 
$\bar{u}^\mu = d\bar{x}^\mu/d\tau$, where $\tau$ is its proper time. This test observer is accelerated; in fact, nongravitational forces are necessary to keep the observer spatially at rest and prevent it from falling into the black hole. The observer carries an adapted nonrotating orthonormal tetrad frame $e^{\mu}{}_{\hat \alpha}$, where $e^{\mu}{}_{\hat 0} = \bar{u}^\mu$,
\begin{equation}\label{K2}
e_{\hat 0}  = \left(\frac{r^2 + a^2}{\Delta}\right)^{1/2}\,\partial_{\tilde t}\,, \qquad  e_{\hat 1}  = \left(\frac{\Delta}{r^2 + a^2}\right)^{1/2}\,\partial_r\,
\end{equation}
and $(e_{\hat 2}, e_{\hat 3})$ are unit vectors in the plane orthogonal to the rotation axis. The rotational symmetry of the Kerr field about its axis leads to enormous simplification of the general analysis presented in~\cite{Bini:2017uax, Mashhoon:2020tha}. Moreover, the observer's acceleration 
$De^{\mu}{}_{\hat 0}/d\tau = \mathbf{A}^\mu = A^{\hat \alpha}\,e^{\mu}{}_{\hat \alpha}$ is given by
\begin{equation}\label{K3}
(A_{\hat 0}, A_{\hat 1}, A_{\hat 2}, A_{\hat 3})  = (0, A, 0, 0)\,, \qquad  A = \frac{M(r^2-a^2)}{(r^2 + a^2)^{3/2}\,(r^2-2Mr+ a^2)^{1/2}}\,.
\end{equation}

For measurement purposes, the reference observer can in principle set up a quasi-inertial Fermi normal coordinate system based on the local nonrotating tetrad frame along its world line; see~\cite{TLC, Synge, bm77, Belinski:2020dmz, Hoegl:2020hif} and the references cited therein. Imagine an arbitrary reference observer with an adapted nonrotating orthonormal tetrad frame $e^{\mu}{}_{\hat \alpha}$. The observer follows a timelike world line $\bar{x}^\mu(\tau)$ in an arbitrary gravitational field. The observer's local temporal axis is the timelike unit vector $e^{\mu}{}_{\hat 0} = d\bar{x}^\mu/d\tau = \bar{u}^\mu$, while $e^{\mu}{}_{\hat i}$, $i = 1, 2, 3$, are orthogonal unit spacelike gyro axes that form the observer's local spatial frame.  Let us consider the class of spacelike geodesics orthogonal to the reference world line at each event $Q(\tau)$ along $\bar{x}^\mu(\tau)$. In this way, we have a local spacelike hypersurface at $\tau$. Suppose $P$ is an event with spacetime coordinates $x^\mu$ on this local spacelike hypersurface such that there is a unique spacelike geodesic of proper length $\sigma$ that connects $P$ to $Q$. The unit spacelike vector $\xi^\mu (\tau)$ is tangent to this spacelike geodesic segment at $Q$. We define the Fermi coordinates of $P$ to be $X^{\hat \mu} = (T, X^{\hat i})$, $X^{\hat i} = (X, Y, Z)$, where
\begin{equation}\label{K4}
T = \tau\,, \qquad  X^{\hat i} = \sigma\,\xi^\mu e_{\mu}{}^{\hat i}\,.
\end{equation}
It follows that the Fermi coordinate time $T$ is the proper time of the reference observer that is located at $X=Y=Z=0$; that is, the reference observer is permanently at rest and occupies the spatial origin of Fermi normal coordinates. The reference observer is furthermore locally inertial in the Fermi system by Einstein's principle of equivalence, but acceleration and curvature terms appear in the metric  away from the location of the fiducial observer at the origin of the spatial Fermi coordinates. In the case under consideration, Fermi coordinates $X^{\hat \mu} = (T, X, Y, Z)$ for Kerr spacetime are such that $X$ denotes the radial coordinate along the Kerr symmetry axis, while $Y$ and $Z$ are the two transverse coordinates.  The metric in Fermi coordinates is given by
\begin{equation}\label{K5}
- ds^2 = g_{\hat \mu \hat \nu}\,dX^{\hat \mu} \,dX^{\hat \nu}\,,
\end{equation}
where the components of the Fermi metric tensor can be expressed as power series in $X$, $Y$ and $Z$ starting from the corresponding components of the Minkowski metric tensor $\eta_{\hat \mu \hat \nu}$, namely,  
\begin{equation}\label{K6}
g_{\hat 0 \hat 0} = - (1+A_{\hat k}\,X^{\hat k})^2 - R_{\hat 0 \hat i \hat 0 \hat j}\,X^{\hat i} X^{\hat j} + O(|\mathbf{X}|^3)\,,
\end{equation}
\begin{equation}\label{K7}
g_{\hat 0 \hat i} =  - \frac{2}{3}\,R_{\hat 0 \hat j \hat i \hat k}\,X^{\hat j} X^{\hat k} + O(|\mathbf{X}|^3)\,
\end{equation}
and
\begin{equation}\label{K8}
g_{\hat i \hat j} = \delta_{\hat i \hat j} - \frac{1}{3}\,R_{\hat i \hat k \hat j \hat l}\,X^{\hat k} X^{\hat l} + O(|\mathbf{X}|^3)\,.
\end{equation}
The components of the curvature tensor (and its covariant derivatives) as measured by the reference observer appear in the coefficients of these power series. For instance, $R_{\hat \alpha \hat \beta \hat \gamma \hat \delta}$ is the projection of the curvature tensor on the tetrad frame of the reference observer, namely, $R_{\mu \nu \rho \sigma}e^{\mu}{}_{\hat \alpha}e^{\nu}{}_{\hat \beta}e^{\rho}{}_{\hat \gamma}e^{\sigma}{}_{\hat \delta}$, which, in a Ricci-flat region of spacetime, can in general be expressed via the symmetries of the curvature tensor in the standard manner as a  $6\times 6$ matrix of the form
\begin{equation}\label{Ka}
\left[
\begin{array}{cc}
\mathbb{E} & \mathbb{H}\cr
\mathbb{H} & -\mathbb{E}\cr 
\end{array}
\right]\,.
\end{equation}
Here, the symmetric and traceless $3\times 3$ matrices $\mathbb{E}$ and $\mathbb{H}$ represent the measured gravitoelectric and  gravitomagnetic components of Weyl curvature, respectively.
These have been worked out in~\cite{Mashhoon:2020tha} for the general case of fiducial observers at rest in the exterior Kerr spacetime.

Fermi coordinates are properly defined within a coordinate patch whose boundaries are given by the standard GR admissibility conditions. The spacetime coordinates are admissible if the corresponding metric with $(-, +, +, +)$ signature is negative definite. That is, we require the metric tensor in matrix form $(g_{\hat \mu \hat \nu})$ to be negative definite; equivalently,  all of  its principal minors must be  negative~\cite{Bini:2012ht, LL}. This means, in particular, that $g_{\hat 0 \hat 0} < 0$ within the Fermi coordinate patch. Therefore, if $g_{\hat 0 \hat 0}$ vanishes,  the corresponding domain where $g_{\hat 0 \hat 0} = 0$ must be part of the boundary of the Fermi coordinate system. 

Timelike and null geodesic motions in the exterior Kerr spacetime have been extensively investigated~\cite{Chandra, BON}. These motions basically refer to the static inertial observers that are at rest at infinity in this asymptotically flat spacetime. However, in connection with motions relative to the Kerr source, we are interested in the timelike motion of free test particles in the Fermi coordinate system established around the world line of our static fiducial observer. The rest mass of such a free test particle is a nonzero constant of its motion; however, we will not explicitly consider the rest mass of the particle further since it is not necessary for the purposes of the present work. The geodesic equation in the Fermi frame can then be obtained from the action principle
\begin{equation}\label{K9}
\delta \int  ds =  \delta \int  \mathcal{L} \,dT = 0\,, \qquad \mathcal{L} :=  \frac{ds}{dT}\,, 
\end{equation}
where the Lagrangian is given by 
\begin{equation}\label{K10}
  \mathcal{L} = ( - g_{\hat 0 \hat 0}- 2 \,g_{\hat 0 \hat i}\, V^{\hat i} - g_{\hat i \hat j}\,V^{\hat i}V^{\hat j})^{1/2}\,, \qquad V^{\hat i} := \frac{dX^{\hat i}}{dT}\,.
\end{equation}
The corresponding  Euler-Lagrange equation is equivalent to the reduced geodesic equation in Fermi coordinates, namely, 
\begin{equation}\label{K11}
\frac{d^2 X^{\hat i}}{dT^2}+\left(\Gamma^{\hat i}_{\hat \alpha \hat \beta}-\Gamma^{\hat 0}_{\hat \alpha \hat \beta}V^{\hat i} \right) \frac{dX^{\hat \alpha}}{dT}\frac{dX^{\hat \beta}}{dT}=0\,,
\end{equation} 
which is a generalization of the Jacobi equation in Fermi coordinates. A detailed description of the Lagrangian and Hamiltonian aspects of such generalized deviation equations is 
contained in~\cite{Chicone:2002kb}. Furthermore, simple generalizations are possible; for instance, we can consider accelerated motion of test particles in the Fermi coordinate system as well.  We must ensure that Eq.~\eqref{K11} refers to timelike motion of the free test particle; to this end, let us define its 4-velocity in the Fermi system 
\begin{equation}\label{K12}
U^{\hat \mu} := \frac{dX^{\hat \mu}}{ds} = \Gamma (1, \mathbf{V})\,, \qquad \mathbf{V} = \frac{d\mathbf{X}}{dT}\,,
\end{equation}
where the Lorentz factor $\Gamma = dT/ds$ can be determined via $U^{\hat \mu}\,U_{\hat \mu} = -1$. Hence, the timelike condition is given by
\begin{equation}\label{K13}
\frac{1}{\Gamma^2} = - g_{\hat 0 \hat 0} - 2\,g_{\hat 0 \hat i}\, V^{\hat i} - g_{\hat i \hat j}\, V^{\hat i} \,V^{\hat j} > 0\,.
\end{equation}
Here, it is important to note that $\mathbf{V}$ is a \emph{ Fermi coordinate velocity}; however, we must have $|\mathbf{V}| <1$ at the location of the reference observer ($\mathbf{X} = 0$). 
It is important to compare and contrast these exact general relativistic tidal equations with the corresponding Newtonian tidal equations presented in Appendix A. 

\section{Exact Tidal Dynamics}

The construction of exact Fermi coordinate systems has thus far been possible only for a few simple spatially homogeneous spacetimes~\cite{Chicone:2005vn, Klein:2009gz}. It was shown explicitly in~\cite{Chicone:2005vn} that it is not possible to express exact Fermi coordinates in Schwarzschild spacetime using only the elementary functions of mathematical physics. In this connection, 
see also~\cite{BGJ} and the references cited therein.
 The same argument extends to the exterior Kerr spacetime as well.  The natural option is to  expand the coordinate transformation from the standard Kerr coordinates to the corresponding Fermi coordinates in infinite series.  Unfortunately, obtaining the coefficients of the power series expansion of the corresponding Fermi metric is nontrivial~\cite{Bini:2005xt, Klein:2007xj}; in fact, this general computation would seem to require a method that is not at present readily available.   In our approximate treatment, we therefore use the expansion of the Kerr metric in Fermi coordinates to second order in the Fermi spatial coordinates  $(X, Y, Z)$; hence, the corresponding tidal equations are given to linear order in $(X, Y, Z)$. Higher-order tidal terms have been studied in~\cite{Marzlin:1994ia, Chicone:2005da}, but not for the Kerr metric.

The Fermi coordinates are introduced here about the world line of a fiducial observer \emph{at rest} in space in the background \emph{stationary} Kerr spacetime; therefore,  
the corresponding Fermi metric must be independent of the Fermi temporal coordinate $T$. It follows that in the case under consideration in this paper, the metric coefficients are of the form  $g_{\hat \mu \hat \nu}(X, Y, Z)$ and  $\partial_T$ is a  timelike Killing vector field. The projection of the 4-velocity vector of a geodesic path on a Killing vector field is a constant of the motion. For a free test particle in the Fermi frame, we thus have
\begin{equation}\label{E1}
U \cdot \partial_T = - \mathcal{E}\,, \qquad \mathcal{E} = -\Gamma \left(g_{\hat 0 \hat 0} + g_{\hat 0 \hat i}\,V^{\hat i}\right)\,, 
\end{equation}
where  $\mathcal{E}$ is proportional to the particle's energy. To see this, let us note that if the free test particle starts out from the location of the fiducial observer with initial speed $V_0$, $0 < V_0 < 1$, 
then from Eqs.~\eqref{K13}--\eqref{E1} and $g_{\hat \mu \hat \nu}(\mathbf{X}=0) = \eta_{\hat \mu \hat \nu}$ we have
\begin{equation}\label{E2}
\mathcal{E} = \Gamma (\mathbf{X}=0) = \frac{1}{(1-V_0^2)^{1/2}}\,.
\end{equation}  

To monitor the motion of the free test particle within the Fermi frame, we imagine the class of observers that are all spatially at rest within the Fermi patch. The fiducial observer is a member of this class of generally accelerated observers that constitute the natural rest frame of the Kerr source. Let $\Lambda^{\hat \mu}{}_{\tilde \alpha}(\mathbf{X})$ be the adapted tetrad frame field of these observers; then, 
$\Lambda^{\hat \mu}{}_{\tilde 0} = (-g_{\hat 0 \hat 0})^{-1/2}\,\delta^{\hat \mu}_{\tilde 0}$ and the 4-velocity vector of the free test particle as measured by this class of observers is given by
\begin{equation}\label{E3}
U_{\tilde \alpha}  = U_{\hat \mu}\,\Lambda^{\hat \mu}{}_{\tilde \alpha}\,.
\end{equation} 
In particular, we can write
\begin{equation}\label{E4}
U^{\tilde \alpha} = \tilde{\Gamma} (1, \tilde{\mathbf{V}})\,, \qquad \tilde{\Gamma} = U^{\tilde 0} = - (-g_{\hat 0 \hat 0})^{-1/2}\,g_{\hat 0 \hat \mu}\,U^{\hat \mu}\,.
\end{equation} 
Equation~\eqref{E1} then implies 
\begin{equation}\label{E5}
\tilde{\Gamma}  = \frac{\mathcal{E}}{(- g_{\hat 0 \hat 0})^{1/2}}\,.
\end{equation}
As measured by the class of static observers in the Fermi frame, a free test particle that comes very close to the two-dimensional boundary of the Fermi coordinate system where 
$g_{\hat 0 \hat 0}(\mathbf{X}) = 0$ has a Lorentz factor that approaches infinity and a speed that is very close to the speed of light. 

We have placed the fiducial observer on the axis of symmetry in direct connection with the motion of jet particles on or near the Kerr rotation axis, which is the $X$ axis in the Fermi frame. Moreover,  the Kerr metric is axially symmetric about its axis of rotation and  $\partial_{\phi}$ is a Killing vector field. The position of the reference observer as well as the construction of a Fermi system about its world line respects this axial symmetry of the underlying geometry. Hence, there must be axial symmetry about the $X$ axis in our Fermi system. In particular, introducing spherical polar coordinates via $X = \rho\,\cos \vartheta$, $Y = \rho\,\sin \vartheta \cos\varphi$ and $Z = \rho\,\sin \vartheta \sin\varphi$,  the exact Fermi metric tensor (expressed in infinite series) must be independent of $\varphi$. It is straightforward to verify this circumstance explicitly using the approximate Fermi metric tensor employed in this paper. These results are compatible with the existence of the Killing vector field $\partial_{\varphi}$ within the Fermi patch.

Azimuthal symmetry about the $X$ axis indicates that jet particles  moving outward along the Kerr rotation axis approach the 
$g_{\hat 0 \hat 0}(\mathbf{X}) = 0$ boundary of the Fermi patch \emph{perpendicularly}. Consider the surfaces of constant $g_{\hat 0 \hat 0}(\mathbf{X})$ and their normal vectors
\begin{equation}\label{E6}
N_{\hat \mu}  = \frac{\partial g_{\hat 0 \hat 0}(\mathbf{X})}{\partial X^{\hat \mu}}|_{g_{\hat 0 \hat 0}(\mathbf{X})={\rm constant}}\,, \qquad N_{\hat 0}  = 0\,.
\end{equation}
At the position of the fiducial observer $\mathbf{X} = 0$ and $g_{\hat 0 \hat 0} = -1$,  and we find from Eqs.~\eqref{K6}--\eqref{K8} that $N^{\hat \mu} = (0, 1, 0, 0)$ coincides with the $X$ axis. Moreover, in the approximation where $O(|\mathbf{X}|^3)$ terms are dropped in Eqs.~\eqref{K6}--\eqref{K8}, the $X$ axis is again orthogonal to the surfaces of constant $g_{\hat 0 \hat 0}(\mathbf{X})$. We expect that this should indeed be the case for the exact Fermi system based on the axial symmetry of the configuration under consideration in this paper. 

What are the conditions that must be satisfied for a realistic jet scenario? Naturally, to get away from the gravitational attraction of the Kerr source and join the jet current, the free test particle must have an initial speed that is above a certain threshold. This circumstance is in conformity with the notion of escape velocity that is well known in Newtonian gravitation. The rest of this paper is devoted to studying the motion of free test particles using the approximate tidal equations based on the Fermi coordinate system where in Eqs.~\eqref{K6}--\eqref{K8} third and higher-order terms in the metric tensor are neglected. However, before we move on to the approximate Fermi system, let us mention an important consequence of Eq.~\eqref{K11}. 

The dynamical system~\eqref{K11} has a rest point given by $\Gamma^{\hat i}_{\hat 0 \hat 0}(X, Y, Z) = 0$, which, from the definition of Christoffel symbols and  the temporal independence of the Fermi metric under consideration, reduces to $g_{\hat 0 \hat 0, \hat i} = 0$. Physically, this means the influence of gravity is zero at a rest point. In the Kerr metric, the influence of gravity decreases as one gets far away from the source and vanishes at spatial infinity. The Fermi coordinate patch is local and does not extend to infinity; hence, there should be no rest point in exact Fermi coordinates. The absence of a rest point is also evident in the Newtonian tidal equation. Thus, a reasonable conjecture is that   $g_{\hat 0 \hat 0, \hat i}$  does not vanish within the Fermi coordinate patch.  The  approximate first-order tidal equations in Fermi coordinates  for the Kerr field do have an \emph{unstable} rest point, a fact that is compatible with the conjecture.   We will prove this elementary result after a more detailed description of these tidal equations and their nondimensionalization. Further discussion of this issue is contained in the following sections.

\section{Approximate Tidal Equations}

Keeping terms only up to  second order in the infinite series in Eqs.~\eqref{K6}--\eqref{K8}, Eq.~\eqref{K11} takes the form
 \begin{eqnarray}\label{S1}
&&\frac{d^2X^{\hat i}}{dT^2}+A_{\hat i}+(R_{\hat 0 \hat i \hat 0 \hat l}+A_{\hat i}\,A_{\hat l})X^{\hat l}-\frac{dA_{\hat l}}{dT}\,X^{\hat l} V^{\hat i} \nonumber\\
&& \qquad\,{}{}-2\,R_{\hat 0 \hat l \hat i \hat j}V^{\hat j}X^{\hat l}-2\,[A_{\hat j}+(R_{\hat 0 \hat j \hat 0 \hat l}-A_{\hat j}A_{\hat l})\,X^{\hat l}]V^{\hat i}V^{\hat j} \nonumber\\
&& \qquad\,{}{} -\frac{2}{3} \left(R_{\hat i \hat j \hat k \hat l}+R_{\hat 0 \hat j \hat k \hat l}V^{\hat i} \right) X^{\hat l}V^{\hat j}V^{\hat k} = 0\,.
\end{eqnarray}
This is the generalized Jacobi equation~\cite{Chicone:2002kb}, which is the reduced geodesic equation in the Fermi system expanded to first-order in the spatial variables. The corresponding approximate timelike condition can be expressed as
\begin{align}\label{S2}
\frac{1}{\Gamma^2} = {}&  (1+A_{\hat i}\,X^{\hat i})^2 - \delta_{\hat i \hat j}\,V^{\hat i}\,V^{\hat j} + R_{\hat 0 \hat i \hat 0 \hat j}\,X^{\hat i}\,X^{\hat j}  \\  \nonumber
 & +\frac{4}{3}\, R_{\hat 0 \hat j \hat i \hat k}V^{\hat i}  X^{\hat j}\,X^{\hat k}+ \frac{1}{3}\,R_{\hat i \hat k \hat j \hat l}\,V^{\hat i}\,V^{\hat j}\,X^{\hat k}\,X^{\hat l} > 0\,.
\end{align}
Let us note that in the approximate system~\eqref{S1}, the rest point occurs at  the solution of the equation
\begin{equation}\label{S3}
A_{\hat i}+(R_{\hat 0 \hat i \hat 0 \hat l}+A_{\hat i}\,A_{\hat l})X^{\hat l} = 0\,.
\end{equation}
However, as mentioned before, the exact tidal equations do not have rest points. 

For the sake of clarity, we briefly digress to explain our approach to approximating tidal force dynamics: We derive consequences of the first-order approximate Eq.~\eqref{S1} as though it were the true equation of motion. Some of the consequences of this equation generally belong to the actual system (e.g., the threshold speed) and some do not (e.g., the rest point). The exact system, given in terms of infinite series in powers of $X$, $Y$ and $Z$, is not explicitly known; therefore, we resort instead to the investigation of the first-order system~\eqref{S1}. Limitations of the approximation are emphasized throughout.  

Specializing these general equations to the case of the fiducial observer fixed at the Boyer-Lindquist radial coordinate $r$ along the rotation axis of the Kerr source, we note that
 the relevant acceleration parameter A is given by Eq.~\eqref{K3}, while the gravitoelectric and the gravitomagnetic curvature components are given by $\mathbb{E}$ = diag$(-2E, E, E)$ and $\mathbb{H}$ = diag$(-2H, H, H)$, where~\cite{Bini:2017uax, Mashhoon:2020tha}
\begin{equation}\label{S4}
E := \frac{Mr(r^2-3a^2)}{(r^2 + a^2)^3}\,, \qquad  H := -\frac{Ma(3r^2-a^2)}{(r^2 + a^2)^3}\,.
\end{equation}
The tidal equations of motion to linear order in $(X, Y, Z)$ are then given by  
\begin{align}\label{S5}
\nonumber  \frac{d^2X}{dT^2} {}&+ (A -2E\,X)\,(1-2\dot{X}^2) +A^2\,X\,(1+2\dot{X}^2)  \\   
 &-\tfrac{2}{3}\,E\, [X(\dot{Y}^2 + \dot{Z}^2) + 2\,\dot{X}(Y\dot{Y} + Z\dot{Z})] + 2\,H (1-\dot{X}^2)(Y\,\dot{Z} - Z\,\dot{Y}) = 0\,,
\end{align}
\begin{align}\label{S6}
\nonumber  \frac{d^2Y}{dT^2} {}&+ E\,Y\,(1-2\dot{Y}^2) -2\,[A- (A^2 + \tfrac{7}{3}\,E)\,X]\,\dot{X}\,\dot{Y} -\tfrac{2}{3}\,E\, [Y(\dot{X}^2 - 2\,\dot{Z}^2) + 5\,Z\,\dot{Y}\,\dot{Z}] \\   
 & + 2\,H\, [2\,X\,\dot{Z} + Z\,\dot{X} - \dot{X}\,\dot{Y}(Y\,\dot{Z} - Z\,\dot{Y})] = 0\,,
\end{align}
\begin{align}\label{S7}
\nonumber  \frac{d^2Z}{dT^2} {}&+ E\,Z\,(1-2\dot{Z}^2) -2\,[A- (A^2 + \tfrac{7}{3}\,E)\,X]\,\dot{X}\,\dot{Z} -\tfrac{2}{3}\,E\, [Z(\dot{X}^2 - 2\,\dot{Y}^2) + 5\,Y\,\dot{Y}\,\dot{Z}] \\   
 & - 2\,H\, [2\,X\,\dot{Y} + Y\,\dot{X} + \dot{X}\,\dot{Z}(Y\,\dot{Z} - Z\,\dot{Y})] = 0\,.
\end{align}
To these equations we must add the condition that the tidal motion is timelike, namely, 
\begin{align}\label{S8}
\nonumber  \frac{1}{\Gamma^2}{}& = (1+ A\,X)^2 - (\dot{X}^2 + \dot{Y}^2+ \dot{Z}^2)- E\, (2\,X^2-Y^2-Z^2) \\
 & +\tfrac{1}{3}\,E\,[ 2\,(Y\,\dot{Z} - Z\,\dot{Y})^2 - (X\,\dot{Z} - Z\,\dot{X})^2 - (X\,\dot{Y} - Y\,\dot{X})^2] + 4\,H X\, (Y\,\dot{Z} - Z\,\dot{Y}) > 0\,.
\end{align}

A timelike geodesic in the Kerr gravitational field has three integrals of motion. Kerr spacetime has two Killing vector fields, leading to the energy and orbital angular momentum constants of the motion,  and a Killing tensor field that leads to Carter's constant.  To take full advantage of these symmetries in the Fermi system, exact Fermi coordinates are required.  Unfortunately, approximations derived from  truncations of series expansions do not necessarily respect first integrals. Nevertheless,  the azimuthal symmetry of Kerr geometry is reflected in our approximate autonomous equations. That is, keeping $(X, \dot{X})$ unchanged at a given time $T$, one can show that the $(Y, Z)$ system of equations is invariant under a rotation by a constant angle about the axis of symmetry.  Moreover, inspection of the $(Y, Z)$ system reveals that $Y(T) = 0$ and $Z(T) = 0$ are not separately possible solutions so long as $H \ne 0$. We note from Eq.~\eqref{S4} that $H = 0$ if either $a = 0$, so that the source is spherically symmetric and described by the Schwarzschild spacetime, or $r = a /\sqrt{3}$. 

Henceforth, we generally assume for the sake of definiteness that in Kerr spacetime with $a \ne 0$,
\begin{equation}\label{S9a}
r \gg M\,, \qquad  r \gg a\,,
\end{equation} 
which are consistent with observable motions of energetic particles relative to ambient astrophysical environments of interest. Therefore, $E > 0$ and $H < 0$.

\subsection{Dimensionless Equations}

It proves convenient to define 
\begin{equation}\label{S9}
(t, x, y, z) := \frac{1}{r} (T, X, Y, Z)\,, \quad (u, v, w) := (\dot{X}, \dot{Y}, \dot{Z})\,, \quad (\mathbb{U}, \mathbb{V}, \mathbb{W}) := r\,(\ddot{X}, \ddot{Y}, \ddot{Z})
\end{equation} 
and
\begin{equation}\label{S10}
\alpha:= r A\,, \quad \beta := r^2 E\,, \quad \,, \quad \gamma:= -r^2 H\,.
\end{equation}
For  $r > \sqrt{3}\,a$, the system parameters $\alpha$, $\beta$ and $\gamma$ are all positive,  and they are functions of the two quantities $M/r$ and $a/r$ given by
\begin{gather}
\nonumber \alpha = \frac{M}{r}\,\frac{1-a^2/r^2}{(1+a^2/r^2)^{3/2}\,(1 -2M/r + a^2/r^2)^{1/2}},\\
\label{S12}
\beta= \frac{M}{r}\,\frac{1-3a^2/r^2}{(1+a^2/r^2)^3}\,, \qquad \gamma =  \frac{M}{r} \,\frac{a}{r}\,\frac{3-a^2/r^2}{(1+a^2/r^2)^3}\,.
\end{gather}
In the extreme Kerr black hole case $a = M$; and, on the axis of symmetry,  $M < r<\infty$. Thus, in the exterior of extreme Kerr black hole along the axis $r > a$ and $\alpha > 0$.  Similarly  $\gamma>0$, but $\beta$ could become negative or zero.

For $r \gg M$ and $r \gg a$, the system parameters may be expanded in series
\begin{gather}
\nonumber \alpha = \frac{M}{r} \left(1 + \frac{M}{r} - 3\,\frac{a^2}{r^2} + \cdots \right)\,, \qquad \beta = \frac{M}{r} \left(1 - 6\,\frac{a^2}{r^2} + \cdots \right)\,,\\
\label{S12a}
\gamma = 3\,\frac{M}{r}\frac{a}{r} \left(1 - \frac{10}{3}\,\frac{a^2}{r^2} + \cdots \right)\,,
\end{gather}
where the neglected higher-order terms are expressible in powers of the small quantities $M/r$ and $a^2/r^2$. 

Using the new dimensionless quantities,  the tidal equations are 
\begin{align} \label{S12b}
(dx/dt, dy/dt, dz/dt) = (u, v, w), \qquad (du/dt, dv/dt, dw/dt)= (\mathbb{U}, \mathbb{V}, \mathbb{W}), 
\end{align}
where
\begin{align}
\nonumber  \mathbb{U} ={}& - (\alpha -2\beta\,x)\,(1-2 u^2) -\alpha^2\,x\,(1+2 u^2)  \\   
\label{S13} &+\tfrac{2}{3}\,\beta\, [x(v^2 + w^2) + 2\,u(y\,v + z\,w)] + 2 \gamma(1-u^2)(yw-zv)\,,\\
\nonumber  \mathbb{V} ={}& -\beta\,y\,(1-2 v^2) +2\,[\alpha- (\alpha^2 + \tfrac{7}{3}\,\beta)\,x]\,u\,v \\
\label{S14} & +\tfrac{2}{3}\,\beta\, [y(u^2 - 2\,w^2) + 5\,z\,v\,w]+ 2 \gamma[2xw+zu-uv(yw-zv)]\,, \\ 
\nonumber  \mathbb{W} ={}& -\beta\,z\,(1-2 w^2) +2\,[\alpha- (\alpha^2 + \tfrac{7}{3}\,\beta)\,x]\,u\,w \\
\label{S15}& +\tfrac{2}{3}\,\beta\, [z(u^2 - 2\,v^2) + 5\,y\,v\,w]- 2\gamma[2xv+yu+uw(yw-zv)]\,.  
\end{align}
The timelike condition is 
\begin{align}\label{S16}
\nonumber \frac{1}{\Gamma^2} = {}& (1+ \alpha\,x)^2 - (u^2 + v^2+ w^2)- \beta\, (2\,x^2-y^2-z^2) \\
 & +\tfrac{1}{3}\,\beta\,[ 2\,(y\,w - z\,v)^2 - (x\,w - z\,u)^2 - (x\,v - y\,u)^2] - 4 \gamma x (yw-zv) > 0\,.
\end{align}

From the symmetry of the configuration under consideration here, we see that for $y = z =0$, the motion can only take place along the radial direction~\cite{Bini:2017uax, Mashhoon:2020tha}.  Further developments in this case in connection with jet motion are contained in the next section. 

By truncating the Lagrangian and the Hamiltonian of the exact tidal equations at second order,  corresponding quantities for the approximate first-order autonomous system~\eqref{S12b}--\eqref{S15} are obtained.  The  truncated Hamiltonian gives an approximate energy only to the extent that motion is limited to a small neighborhood of the spatial origin respecting the first-order approximation in $x$, $y$ and $z$; but, of course, this energy is not a constant of the motion. In fact, first integrals for system~\eqref{S12b}--\eqref{S15} most likely do not exist.

\subsection{Exact Solutions of Approximate Tidal Equations}

The nonlinear autonomous ordinary differential tidal equations produce a local flow in the six-dimensional state space  $(x, y, z, u, v, w)$ where solutions are physically relevant when they remain timelike within a valid Fermi coordinate patch. A complete description of their dynamics does not seem possible.  But the system has a rest point  in agreement with Eq.~\eqref{S3} and some other identifiable exact solutions. 

With $u$, $v$ and $w$ set to zero in the expressions~\eqref{S13}--\eqref{S15} for $\mathbb{U}$, $\mathbb{V}$  and $ \mathbb{W}$, inspection 
reveals the existence of a single rest point with spatial coordinates
\begin{equation}\label{T1}
 x_{\text{rp}} = \frac{\alpha}{2\,\beta-\alpha^2}\,,  \qquad  y_{\text{rp}} = z_{\text{rp}} = 0\,.
\end{equation}
The timelike condition at the rest point reduces to $2 \beta/(2 \beta-\alpha^2 )> 0$ and is satisfied in the case considered here.  For $r \gg M$ and $r \gg a$, the first coordinate of the rest point may be expanded in powers of  $M/r$ and $a/r$;  in fact, 
\begin{equation}\label{T2}
2 \beta - \alpha^2 = 2\,\frac{M}{r} \left(1 - \tfrac{1}{2}\,\frac{M}{r} - 6\,\frac{a^2}{r^2} + \cdots \right)\,, \qquad x_{\rm rp} = \frac{1}{2}\, \left(1 + \tfrac{3}{2}\,\frac{M}{r} + 3\,\frac{a^2}{r^2} + \cdots   \right)\,.
\end{equation}

Linearization at the rest point produces a constant system matrix with distinct eigenvalues: two complex conjugate pairs of pure imaginary eigenvalues and two real eigenvalues symmetrically located about the origin on the real line.  In fact, the eigenvalues are of the form  $\pm R_1$, $\pm i R_2$ and $\pm i R_3$, where $R_j$, $j = 1, 2, 3$, is real and given by
\begin{gather}\label{T3}
\nonumber 
R_1=(2 \beta-\alpha^2)^{1/2}\,, \quad 
R_2 = \beta -4 \alpha \beta \gamma \left\{2 \alpha \gamma -\left[\beta\,(2 \beta-\alpha^2)^2 + 4\alpha^2 \gamma^2\right]^{1/2}\right\}^{-1}\,,\\
R_3 = \beta -4 \alpha \beta \gamma \left\{2 \alpha \gamma +\left[\beta\,(2 \beta-\alpha^2)^2 + 4\alpha^2 \gamma^2\right]^{1/2}\right\}^{-1}\,.
\end{gather}
The presence of a positive eigenvalue ($R_1$) implies the rest point is unstable.  Because there are pure imaginary eigenvalues, the  rest point is not hyperbolic. But by standard results the nonlinear system has a one-dimensional stable manifold, a one-dimensional unstable manifold (responsible for the instability), and a four-dimensional center manifold. On a full measure set where the eigenvalues are free of resonances, there is moreover a pair of two-dimensional sub-center manifolds corresponding to each pair of complex conjugate pure imaginary eigenvalues. If there were corresponding first integrals (as there would be for the exact tidal equations due to the Hamiltonian structure of geodesic flows) each sub-center manifold would be foliated by one-parameter families of periodic solutions as a consequence of Lyapunov's center theorem~\cite{CC}.  Absent first integrals, a higher-order analysis is required to determine the nonlinear flow on the center manifold (see, for example,~\cite{S} for a survey of this methodology).  We note that an eigenvector generating the unstable direction has components
\begin{align}\label{T4}
(\frac{1}{(2 \beta-\alpha^2)^{1/2}},0,0,1,0,0),
\end{align}
which correspond to the radial direction.  It would be remarkable that within the  tidal framework  a free test particle could  have a rest point in the exterior Kerr spacetime as for the approximate equations of motion. But as mentioned previously,  the exact tidal equations do not have a rest point. Thus further analysis of the behavior of the approximate tidal equations in a neighborhood of the spurious rest point is not pursued; numerical experiments suggesting the dynamical behavior of the approximate motion are reported instead in Section~\ref{sec:ne}.

Regarding the physical interpretation of our results, imagine a gravitational source confined within a compact region of space and a set of far away test masses falling radially toward the source. The attractive nature of gravity implies the steady infall of a test mass should continue until the mass is very close to the source. In any case, no \emph{rest point} is allowed. In this description of motion the reference observers are the static inertial observers at spatial infinity. In the present work, the reference observer is instead radially accelerated and stays at rest somewhere on the axis of rotational symmetry of the exterior Kerr spacetime. It is rather surprising however that there should be a rest point for a free test particle in the approximate Fermi coordinate patch established around the fiducial observer. The rest point must be radially unstable in order to avoid physical inconsistency. In connection with this interpretation,  it is important that the existence of the rest point be directly connected with the acceleration of the reference observer; indeed, we note that $x_{\text{rp}}$ is directly proportional to $\alpha = rA$. 

Inspection of Eqs.~\eqref{S12b}--\eqref{S15} reveals the existence of the following additional exact solutions: $(x_1,  y_1,  z_1)$ and $(x_2, y_2, z_2)$, where 
\begin{gather}\label{T5}
\nonumber
x_1=x_0\,,\quad
y_1= y_0 + \sigma \frac{\sqrt{2}}{2}\,(t-t_0)\,,\quad
z_1 = - \sigma \frac{6\sqrt{2}}{5}\frac{\gamma}{\beta} \,x_0\,,\\
u_1=0\,,\quad
v_1= \sigma\,\frac{\sqrt{2}}{2}\,,\quad
w_1=0\,.
\end{gather}
Here, $\sigma = \pm 1$,  $t_0$ and  $y_0$ are arbitrary constants, and
\begin{align}\label{T6}
x_0= \frac{15 \alpha\beta}{35 \beta^2 -15 \alpha^2\beta +36 \gamma^2}\,.
\end{align}
For $r \gg M$ and $r \gg a$, we find
\begin{equation}\label{T7}
x_0 = \frac{3}{7}\, \left(1 + \frac{10}{7}\,\frac{M}{r} + \cdots   \right)\,.
\end{equation}
For $\gamma = 0$, the rotation is turned off and this exact solution reduces to the one investigated in detail before in the context of Schwarzschild spacetime~\cite{Mashhoon:2020tha} and shown to be unstable.  Numerical experiments demonstrate that the solution is unstable  for $\gamma \ne 0$ as well.

Similarly, 
\begin{gather}\label{T8}
\nonumber
x_2=x_0\,,\quad
y_2= \sigma'\, \frac{6\sqrt{2}}{5}\frac{\gamma}{\beta} x_0\,,\quad
z_2 = z_0 + \sigma'\, \frac{\sqrt{2}}{2}\,(t-t_0)\,,\\
u_2=0\,,\quad
v_2=0\,,\quad
w_2=\sigma'\, \frac{\sqrt{2}}{2},
\end{gather}
where $\sigma' = \pm 1$ and $t_0$ and  $z_0$ are arbitrary constants. 
Moreover, let us note that $(x_3,  y_3,  z_3)$, where
\begin{gather}\label{T9}
\nonumber
x_3=x_0\,,\quad
y_3= y_1 \cos\Theta + y_2 \sin\Theta\,,\quad
z_3 = z_1\cos\Theta + z_2 \sin\Theta\,,\\
u_3=0\,,\quad
v_3= \sigma\,\frac{\sqrt{2}}{2} \cos\Theta\,,\quad
w_3=\sigma'\, \frac{\sqrt{2}}{2} \sin\Theta\,,
\end{gather}
is an exact solution of Eqs.~\eqref{S12b}--\eqref{S15} provided
\begin{equation}\label{T10}
(\sigma' y_0 - \sigma z_0)\sin\Theta \cos\Theta = 0\,.
\end{equation}
Here, $\Theta$ is an angle such that $0\le \Theta < 2\pi$. For $\Theta = 0$ or $\pi$, this third solution reduces to the first one, while for  $\Theta = \pi/2$ or $3\pi/2$, 
it reduces to the second one; otherwise, we have a new exact solution with $\sigma' y_0 = \sigma z_0$. 

For this set of exact solutions, the timelike condition is satisfied; furthermore, 
numerical experiments indicate that these solutions are unstable. No clear physical interpretation of these solutions has been possible thus far.

We now turn to a more detailed examination of the implications of the approximate tidal equations.

\section{Approximate Boundary of Fermi Coordinates}

With the metric signature adopted in this paper, the standard admissibility conditions for the Fermi coordinates require that the principal minors of the matrix $(g_{\hat \mu \hat \nu})$ be negative. For the background exterior Kerr spacetime, the components of the Fermi metric up to second order in $(x, y, z)$ are given by
\begin{equation} \label{F1}
g_{\hat 0 \hat 0} = - (1 + \alpha\,x)^2 + \beta (2 x^2 - y^2 - z^2)\,,
\end{equation}
\begin{equation} \label{F1a}
g_{\hat 0 \hat 1} = 0\,,\qquad g_{\hat 0 \hat 2} = -2\gamma\,xz\,, \qquad  g_{\hat 0 \hat 3} = 2\gamma\,xy\,,
\end{equation}
\beq \label{F2}
g_{\hat 1 \hat 1} = 1 + \tfrac{1}{3} \beta\, (y^2+z^2)\,,\quad g_{\hat 1 \hat 2} = -\tfrac{1}{3} \beta\, xy\,,\quad g_{\hat 1 \hat 3} = -\tfrac{1}{3} \beta\, xz\,
\eeq
and
\beq \label{F3}
g_{\hat 2 \hat 2} = 1 + \tfrac{1}{3} \beta\, (x^2 - 2\,z^2)\,,\quad g_{\hat 2 \hat 3} = \tfrac{2}{3} \beta\, yz\,, \quad g_{\hat 3 \hat 3} = 1 + \tfrac{1}{3} \beta\, (x^2 - 2\,y^2)\,.
\eeq

Let us note that the approximate Fermi metric can be written as
\begin{equation}\label{Fa}
g_{\hat \mu \hat \nu}  = \eta_{\hat \mu \hat \nu} - 2\, \alpha\,x \,\eta_{\hat \mu \hat 0}\,\eta_{\hat \nu \hat 0}+ h_{\hat \mu \hat \nu}\,,
\end{equation} 
where  $h_{\hat \mu \hat \nu}$ is of second order in $(x, y, z)$. Furthermore,  the corresponding inverse Fermi metric is given by
\begin{equation}\label{Fb}
g^{\hat \mu \hat \nu}  = \eta^{\hat \mu \hat \nu} + (2\, \alpha\,x - 4 \, \alpha^2\,x^2) \,\delta^{\hat \mu}_{\hat 0}\,\delta^{\hat \nu}_{\hat 0} - h^{\hat \mu \hat \nu}\,, \quad g^{\hat \mu \hat \nu}\,g_{\hat \nu \hat \rho}= \delta^{\hat \mu}_{\hat \rho}\,,
\end{equation} 
where  the indices of the second-order perturbation $h_{\hat \mu \hat \nu}$ are raised and lowered via the Minkowski metric tensor.  For instance, 
\begin{equation} \label{Fc}
h^{\hat 0 \hat 0} = h_{\hat 0 \hat 0} = (2\beta- \alpha^2)\,x^2 - \beta  (y^2 + z^2)\,, 
\end{equation}
etc.

\begin{figure}
\begin{center}
 \includegraphics[width =7cm]{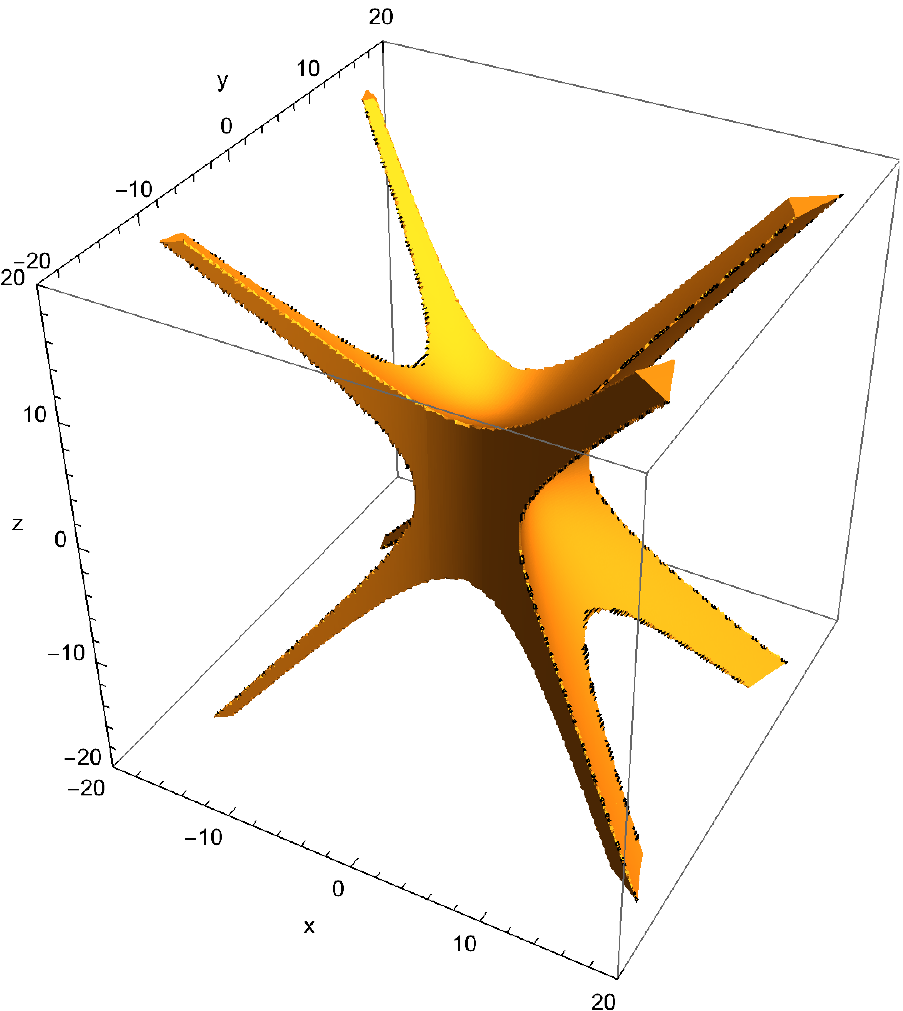}\quad \includegraphics[width =7cm]{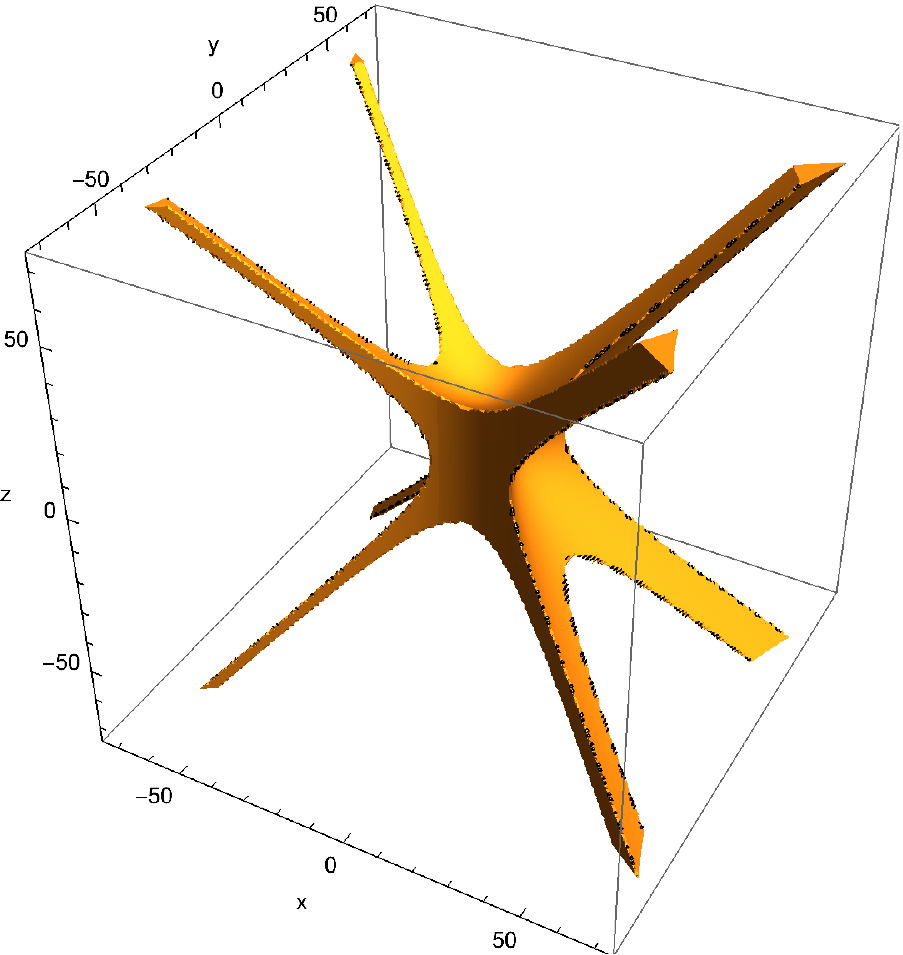}
\end{center}
\caption{The figures are numerical approximations of the region in space defined by the Fermi metric admissibility conditions~\eqref{F4}--\eqref{F6}. The left panel is for $M/r = 0.1$ and $a/r = 0.05$; the right panel is for $M/r = 0.1$ and $a/r = 0.5$.\label{Fig:ferbdy}}
\end{figure}

It follows from a detailed analysis that the admissibility conditions are
\begin{align}
 \label{F4}
g_{\hat 0 \hat 0}  &= - (1 + \alpha\,x)^2 + \beta (2\, x^2 - y^2 - z^2) < 0\,,\\
 \label{F5}
 1& + \tfrac{1}{3}\beta\, (x^2+y^2-z^2) > 0\,,\\
 \label{F6}
g_{\hat 3 \hat 3} &= 1 + \tfrac{1}{3}\beta\, (x^2 - 2\,y^2) > 0\,,
\end{align}
where we have neglected higher-order terms. Three hyperboloids cross and we get the interesting shapes depicted in Fig.~\ref{Fig:ferbdy}. The union of these three hyperboloids is the approximate domain of viability of the Fermi coordinate system under consideration. Let us note here that $g_F = \det(g_{\hat \mu \hat \nu})$ in the Fermi frame is given by
\beq \label{F7}
g_F =  g_{\hat 0 \hat 0} \,[1 + \tfrac{1}{3}\beta\, (x^2 + y^2- z^2)]\,g_{\hat 3 \hat 3}\,,
\eeq
which should be negative in order that Fermi coordinates be admissible. We have assumed here that $\beta > 0$. Moreover,  $g_F < 0$ due to Eqs.~\eqref{F4}--\eqref{F6}.

\subsection{Approximate Admissible Region Along the $x$ Direction} 
 
For tidal motion purely along the $x$ axis (i.e., the radial direction) with $y=z=0$,  the standard admissibility conditions  reduce to the inequality 
\beq \label{F8}
g_{\hat 0 \hat 0}(t, x, 0, 0) = - (1 + \alpha\,x)^2 + 2\beta\, x^2 < 0\,,
\eeq
when  higher-order terms are neglected. Condition~\eqref{F8} for  $r >> M$ and $r >> a$ is easily satisfied, since $\sqrt{2\beta} > \alpha$ in this case and we find
\beq \label{F9}
  x_{-} < x < x_{+}\,, \qquad x_{\pm} =\frac{1}{\pm \sqrt{2\beta} -\alpha}\,. 
\eeq
On the other hand, if  $\sqrt{2\beta} < \alpha$; then, the admissible regimes are $x >  x_{-}$ and $x <  x_{+}$. In case $\sqrt{2\beta} = \alpha$, we have $x > -1/(2\,\alpha)$. 

In connection with the jet problem, it is important to remark that for the outward motion of particles near and along the rotation axis of the Kerr source, the boundary of the Fermi system is part of the hyperboloid given by $g_{\hat 0 \hat 0} = 0$ that crosses the rotation axis orthogonally at $ x_{+}$. 

\section{Approximate Treatment of Motion Along the Rotation Axis}

Motion along the rotation axis of the Kerr solution has been investigated in~\cite{Bini:2017uax, Mashhoon:2020tha}. Here, we pose and  answer some  questions that were not previously considered. 

\begin{figure}[h]
\centerline{ \includegraphics[width =7cm]{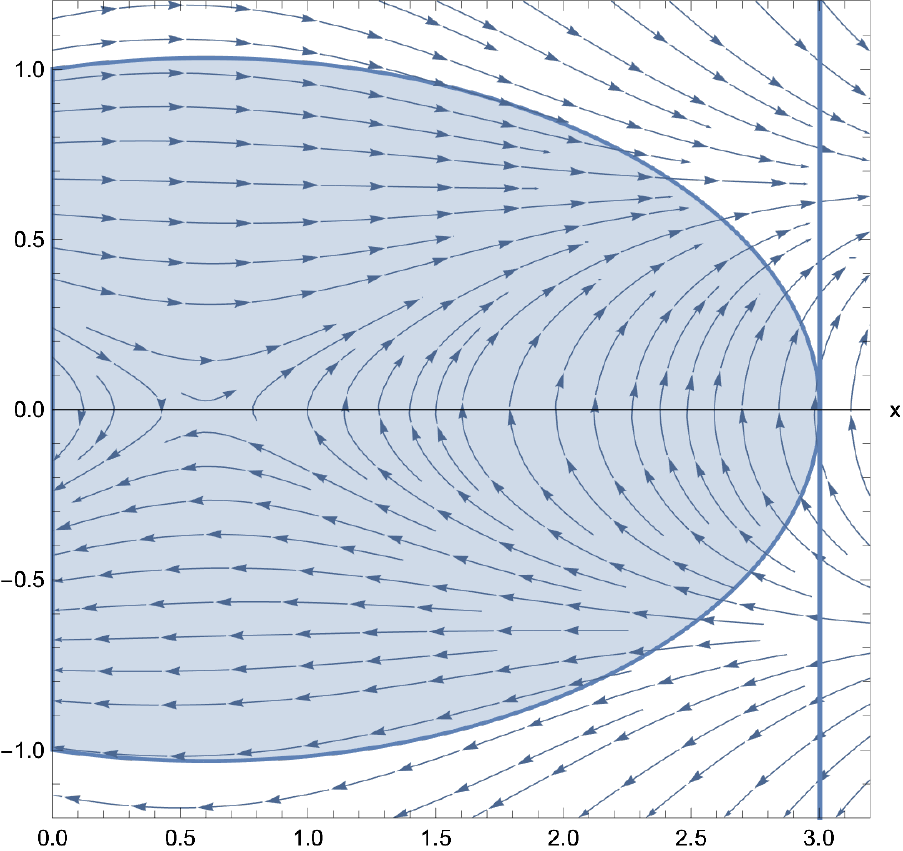}}
\caption{ A typical phase portrait for the moderately rotating ($M = 2a$) Kerr black hole case with  $M/r = 0.1$ and $a/r=0.05$. The figure includes the shaded timelike constraint region and the right-hand boundary of the region of existence of Fermi coordinates (i.e., the vertical line at $x_{+} \approx 3$). We have checked that for the extreme Kerr black hole with $M/r=0.1$ and $a/r=0.1$ the corresponding portrait is essentially the same.   \label{fig:radtide}}
\end{figure}

For $y(t) = z(t) = 0$, tidal motion along the Kerr rotation axis  is governed by the differential equation
\begin{equation}\label{R1}
 \frac{d^2x}{dt^2}+ (\alpha -2\beta\,x)\,(1-2 u^2) + \alpha^2\,x\,(1+2 u^2) = 0\,
\end{equation}
and the timelike constraint
\begin{equation}\label{R2}
(1+ \alpha\,x)^2 - u^2 - 2\,\beta\,x^2 > 0\,.
\end{equation}
In state space, the equation of motion is the first-order system
\begin{align}\label{R2a}
\nonumber \frac{dx}{dt}&=u,\\
\frac{du}{dt}&= -(\alpha -2\beta\,x)(1-2 u^2) - \alpha^2x(1+2 u^2)\,.
\end{align}
Note that the phase portrait $u$ versus $x$ is symmetric across the $x$ axis. 
There is a unique rest point at 
\begin{align}\label{R2b}
x_{\rm rp} = \frac{\alpha}{2\,\beta - \alpha^2}\,,\qquad u_{\rm rp}=0\,.
\end{align}
This exact solution of Eq.~\eqref{R1} corresponds to timelike motion provided $2\beta - \alpha^2 >0$. 
In this case, the rest point is a hyperbolic saddle and represents an unstable equilibrium; moreover, see Eq.~\eqref{T2} and the corresponding discussion presented in Section III.  
Thus, the system has no periodic orbits. Moreover, the  portion of the stable manifold in the upper half-plane ($u>0$) meets the $u$ axis with a unique positive  $u$ coordinate $u_{\rm crit}$. The portion of the unstable manifold of the rest point in the upper half-plane extends to infinity in the horizontal direction. The curve formed by joining these two curves separates solutions starting with $x=0$ and $u>0$ into three  types: For $u< u_{\rm crit}$, the orbit slows down to velocity zero, turns back and falls into the source, for $u=u_{\rm crit}$ the orbit approaches the rest point in infinite time; and for  $u> u_{\rm crit}$ it continues to spatial infinity.  But, all that has been said is limited by the timelike and admissibility constraints, see Fig.~\ref{fig:radtide} for a typical phase portrait for a moderate Kerr black hole with $a/M = 0.5$; the corresponding figure for an extreme Kerr black hole is similar.

The intersection point of the stable manifold of the rest point with the $u$ axis cannot be determined in closed form, but it can be estimated by power series approximation of the stable manifold, numerically, or via the following discussion.

Note that Eq.~\eqref{R1} has the alternate form
\begin{equation}\label{R3}
 \frac{du^2}{dx}+ 4\,[(2\,\beta + \alpha^2)\,x - \alpha] \,u^2 + 2\,[\alpha - (2\,\beta - \alpha^2)\,x] = 0\,.
\end{equation}
It is straightforward to integrate Eq.~\eqref{R3} by means of an integrating factor $\mathcal{I}$, 
\begin{equation}\label{R4}
\mathcal{I}(x) = \exp{[2\,(2\,\beta + \alpha^2)\,x^2 - 4\,\alpha\,x]}\,, 
\end{equation}
where
\begin{equation}\label{R5}
 \frac{d\mathcal{I}}{dx} = 4\,[(2\,\beta + \alpha^2)\,x - \alpha]\,\mathcal{I}\,. 
\end{equation}
With the initial condition that at $t = 0$  a free test particle at $x = 0$ has Fermi coordinate velocity  $u = u_0$, $|u_0| < 1$, Eq.~\eqref{R3} can be integrated once and we find the ``energy" equation
\begin{equation}\label{R6}
\mathcal{I}(x) \,u^2 +\mathcal{W}_{\rm eff}(x) = u_0^2|_{x=0}\,, 
\end{equation}
where $\mathcal{W}_{\rm eff}(x)$ is the effective potential energy 
\begin{equation}\label{R7}
\mathcal{W}_{\rm eff}(x) =2\,\int_0^x [\alpha - (2\,\beta - \alpha^2)\,\xi]\,\mathcal{I}(\xi) \,d\xi\,.
\end{equation}
It follows from the definition of effective potential~\eqref{R7} that this function vanishes at $x = 0$ and has an extremum at $x = x_{\rm rp}$. The path of the free particle $x(t)$ can be simply obtained from integrating Eq.~\eqref{R6}. To clarify the nature of this motion, we note that $\mathcal{I}(x) > 0$; hence, Eq.~\eqref{R6} implies that the motion is confined to the region 
\begin{equation}\label{R8}
\mathcal{W}_{\rm{eff}}(x)\le u_0^2|_{x=0} <1\,.
\end{equation}
\emph{Free test particles that start at $x = 0$ and move radially outward with $u_0^2|_{x=0}$ less than the maximum height of the barrier, i.e., $\mathcal{W}_{\rm eff}(x_{\rm rp})$, have turning points and fall back toward the source, while those with $u_0^2|_{x=0} > \mathcal{W}_{\rm eff}(x_{\rm rp})$ eventually accelerate toward the speed of light with $\Gamma = \infty$}~\cite{Bini:2017uax, Mashhoon:2020tha}.

The interpretation of phase portrait in Fig.~\ref{fig:radtide} in terms of scattering by a potential barrier has thus far involved only outgoing curves that hit the barrier on its left side. It is possible to consider incoming curves in the lower half-plane as well. These curves hit the barrier on its right side, have turning points and turn around and go back out (in the upper half-plane) toward the positive $x$ axis and eventually accelerate to the speed of light by crossing the ellipse, see Fig.~\ref{fig:radtide}.  That is, collisions in the jet could lead to free test particles moving downward toward the source; however, some of these could have turning points and become part of the outgoing jet.  

\begin{figure}[h]
\begin{center}
 \includegraphics[width =10cm]{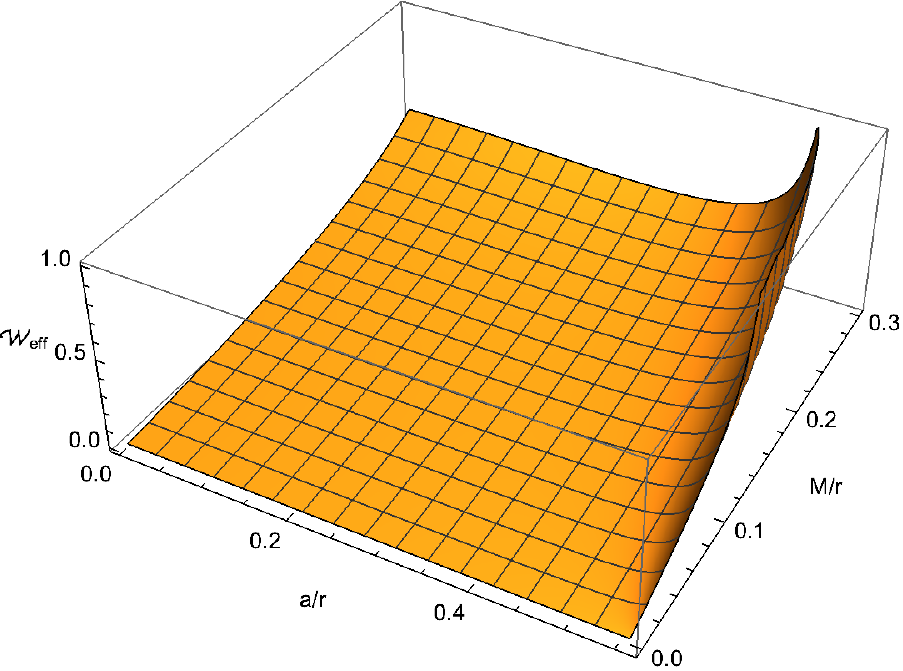}
\end{center}
\caption{A numerical approximation of the effective potential $\mathcal{W}_{\rm{eff}} (x_{\rm rp})$ as a function of $a/r$ and $M/r$ is depicted. The quantity $\mathcal{W}_{\rm{eff}} (x_{\rm rp})$ becomes very large at $(a/r = 0.56, M/r = 0.2)$; however, only the regime $0 \le \mathcal{W}_{\rm{eff}} (x_{\rm rp}) < 1$ is physically meaningful.\label{fig:wsg}}
\end{figure}

\begin{figure}[h]
\begin{center}
 \includegraphics[width =7cm]{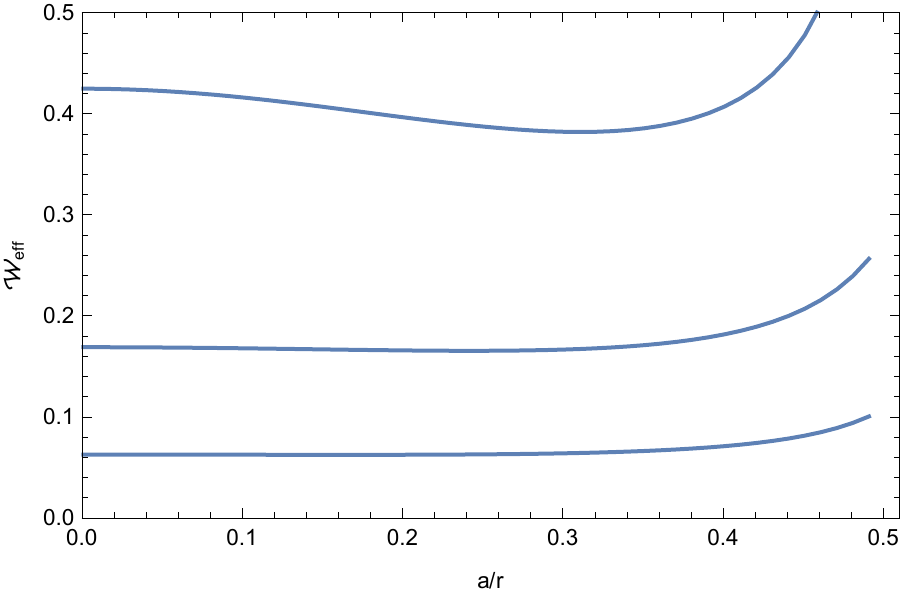}
\end{center}
\caption{The effective potential $\mathcal{W}_{\rm{eff}} (x_{\rm rp})$,  $\mathcal{W}_{\rm{eff}} (x_{\rm rp}) < 1$,  as a function of $a/r$  is depicted for three values of $M/r$: The top graph is for $M/r=0.3$, the middle for $M/r=0.2$, and the bottom for $M/r=0.1$. We note that the threshold speed for tidal acceleration to near the speed of light, $u_{\rm crit}$,  $u_{\rm crit}^2 = \mathcal{W}_{\rm{eff}} (x_{\rm rp}) < 1$, decreases with increasing rotation parameter $a/r$ while the source is a black hole,  namely, $a\le M$. Beyond the black hole regime, $u_{\rm crit}$ increases with $a/M$. This can be clearly seen in the top graph: until $a/r = 0.3$, which occurs at $a =M$, the effective potential at the rest point decreases, but this trend soon stops and the effective potential increases with $a/M > 1$. \label{fig:3gr}}
\end{figure}

Let us recall here that for $r \gg M$ and $r \gg a$, $2\beta -\alpha^2 > 0$. Moreover, assuming that the physical parameters of the configuration under discussion here are such that $2\beta -\alpha^2 > 0$, we find from inspecting the expression for the effective potential energy function $\mathcal{W}_{\rm eff}(x)$ that we have a simple potential barrier here that starts from zero at $x = 0$ and  reaches its maximum value of  $u_{\rm crit}^2$ at $x_{\rm rp}$, where the threshold speed  $u_{\rm crit} > 0$ is given by
\begin{equation}\label{R9}
u_{\rm crit}^2 := \mathcal{W}_{\rm{eff}}(x_{\rm rp}) =2\,\int_0^{x_{\rm rp}} [\alpha - (2\,\beta - \alpha^2)\,\xi]\,e^{[2\,(2\,\beta + \alpha^2)\,\xi^2 - 4\,\alpha\,\xi]} \,d\xi\,.
\end{equation}
The energy Eq.~\eqref{R6} implies that at the rest point $(x, u) = (x_{\rm rp}, 0)$, $u_{\rm crit}^2 := u_0^2|_{x = 0}$. Figs.~\ref{fig:wsg} and~\ref{fig:3gr} explore the nature of the $ \mathcal{W}_{\rm{eff}}(x_{\rm rp})$ as a function of $a/r < 1$ and $M/r<1$ such that $2\beta -\alpha^2 > 0$. In particular, Fig.~\ref{fig:3gr} demonstrates that for fixed $M/r$, the threshold speed decreases with increasing $a/r$ for a black hole, which would make it easier for free test particles to get tidally accelerated along the rotation axis of the black hole and join the jet.

\section{Numerical Experiments\label{sec:ne}} 

\begin{figure}[h]
\begin{center}
 \includegraphics[width =7cm] {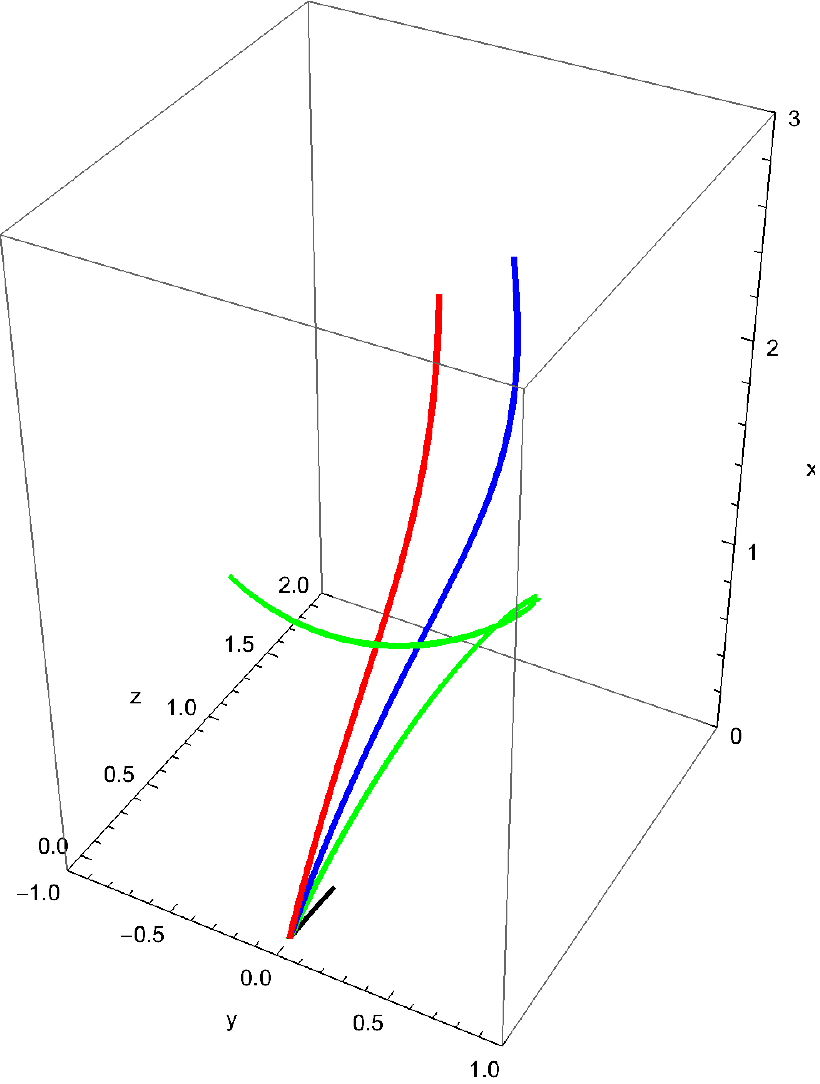}
\end{center}
\caption{Plot of the trajectories of free test particles that all start out from the location of the fiducial observer at the origin of spatial Fermi coordinates with speed $0.5$. The initial velocity vectors are all in the $(x, z)$ plane with deviation angles $\pi/6$, $\pi/4$, $\pi/3$ and $\pi/2$ from the $x$ axis. We integrate Eqs.~\eqref{S13}--\eqref{S15} for an extreme Kerr black hole $(a = M)$ with $M/r = 0.1$. The trajectories twist out of the $(x, z)$ plane due to the gravitomagnetic field of the Kerr source represented by parameter $\gamma \ne 0$. The integration stops when a timelike trajectory turns null, i.e., the timelike condition~\eqref{S16} is violated. This happens for all the trajectories shown within the confines of the Fermi coordinate patch.  \label{fig:5}}
\end{figure}

The dominant influence on a free test particle is the gravitational attraction of the Kerr source. Therefore, all of the trajectories of the particles tend to fall toward the Kerr source unless the initial speed of the particle is above a certain threshold in order to allow the particle to escape the gravitational attraction of the central source. This is true for motion in all forward directions away from the source, as previous work in the case of the Schwarzschild spacetime has demonstrated~\cite{RoMa}. However, the rotation of the Kerr source brings in the gravitomagnetic aspect of the motion (due to the rotation parameter $a$) in addition to the gravitoelectric aspect (due to mass $M$) that is shared with the Schwarzschild source. To lowest order, the gravitomagnetic acceleration is present in Eq.~\eqref{S1} in the term $2\,R_{\hat 0 \hat l \hat i \hat j}V^{\hat j}X^{\hat l}$, which is the analogue of the $\mathbf{v}\times \mathbf{B}$ term familiar from the Lorentz force law. This circumstance is reflected in Eqs.~\eqref{S13}--\eqref{S15} in the terms proportional to $\gamma$.   Thus we expect to see the effect of the gravitomagnetic field on the trajectories of free test particles in this case. This is illustrated here in Fig.~\ref{fig:5}. We note that in the recent GP-B experiment~\cite{Francis1, Francis2},  the non-Newtonian gravitomagnetic field of the Earth predicted by GR due to mass current has been directly measured at about the 19\% level. 

\section{Discussion}

The main purpose of this paper is to present a scenario for the tidal acceleration to near the speed of light of free test particles in an astrophysical jet. We choose a fiducial observer located at  a fixed Boyer-Lindquist radial coordinate $r$ along the rotation axis of an exterior Kerr spacetime and establish a quasi-inertial Fermi normal coordinate system based on the nonrotating orthonormal tetrad frame adapted to the fiducial observer. The Fermi patch acts as an effective local rest frame for the Kerr black hole. From Eq.~\eqref{E5}, we know that  if a free test particle makes it to the edge of Fermi frame where $g_{\hat 0 \hat 0} = 0$, then the particle is moving near the speed of light, since its Lorentz factor as measured by observers at rest in our Fermi frame is given by $\tilde{\Gamma}  = \mathcal{E}(- g_{\hat 0 \hat 0})^{-1/2}$, where $\mathcal{E}$ is a constant. To get to the outer boundary of Fermi system along the jet, the free test particle must have an initial speed above a certain threshold, which is the analog of the escape velocity in the present context. To proceed, we use approximate Fermi coordinates and restrict the motion to the rotation axis of the Kerr black hole; indeed, the metric of the corresponding exact Fermi system would involve infinite power series that are not presently available. Free test particles that start from the location of the fiducial observer (i.e., spatial origin of Fermi coordinates) and go outward (as in a jet) along or near the rotation axis have to cross a potential barrier. Below a threshold speed $u_{\rm crit}$, they fall back towards the black hole, but above the threshold they go forward in the jet direction and, in our approximate treatment, they reach $\Gamma = \infty$ given by an ellipse where timelike motion turns null before reaching the edge of the Fermi frame (the straight vertical line at $x \approx 3$ in Fig.~\ref{fig:radtide}).  We conjecture that in the exact Fermi system, the ellipse changes into the boundary of the Fermi system where $g_{\hat 0 \hat 0} = 0$ in agreement with Eq.~\eqref{E5}. The various considerations contained in our paper are brought together in general support of the jet scenario developed in this work.

\appendix

\section{Newtonian Tidal Equations}

Consider the analogue of our tidal equations for a Newtonian source.  Imagine a Cartesian coordinate system $(\bar{x}, \bar{y}, \bar{z})$ and a spherical mass $M$ at its origin. The Newtonian equation of motion for the $\bar{x}$ component of a free test particle in the exterior of the source is
\begin{equation}\label{A1}
 \frac{d^2 \bar{x}}{d\bar{t}^2} = -\frac{GM}{(\bar{x}^2 + \bar{y}^2 + \bar{z}^2)^{3/2}}\,\bar{x}
\end{equation}
with similar equations for $\bar{y}$ and $\bar{z}$. A fiducial observer is fixed in the exterior region at $(\bar{x}_0, \bar{y}_0, \bar{z}_0)$, where  $r: = (\bar{x}_0^2 + \bar{y}_0^2 + \bar{z}_0^2)^{1/2}$. It is clear that the observer has an outward nongravitational acceleration given by $GM/r^2$. Let $X := \bar{x} -  \bar{x}_0$, $Y := \bar{y} -  \bar{y}_0$, and  $Z = \bar{z} -  \bar{z}_0$.
 Taking advantage of the spherical symmetry of the configuration,  choose the $\bar{x}$ axis such that the fiducial observer lies on it such that $\bar{x}_0= r$ and $\bar{y}_0=\bar{z}_0=0$. Also let   $T := c \,\bar{t}$, where $c$ is the speed of light.   In these coordinates, 
\begin{equation}\label{A2}
 \frac{d^2 X}{dT^2} = -\frac{GM(r+X)}{c^2\mathbb{D}^3}\,, \quad  \frac{d^2 Y}{dT^2} = -\frac{GMY}{c^2\mathbb{D}^3}\,,\quad \frac{d^2 Z}{dT^2} = -\frac{GMZ}{c^2\mathbb{D}^3}\,,
 \end{equation}
where 
\begin{equation}\label{A3}
 \mathbb{D} = [(r+X)^2 + Y^2 + Z^2]^{1/2}\, 
\end{equation}
and $1/c^2$ plays the role of a constant multiplicative factor and has no basic physical significance in the present context; it has been introduced to make the comparison with GR equations more convenient.
This system is equivalent to a Hamiltonian system with Hamiltonian 
\begin{align}\label{A4}
\mathcal{H}(X,Y,Z,U,V,W)=\frac{1}{2} (U^2+V^2+W^2)-\frac{GM}{c^2} [(X+r)^2+Y^2+Z^2]^{-1/2}\,, 
\end{align}
where $U := dX/dT$, etc.

\subsection{Approximate Newtonian Equations}
In analogy with the approximation of the GR tidal equations,  expansion of the Newtonian system~\eqref {A2} to first-order in the space variables produces the approximate system
\begin{equation}\label{A5}
 \frac{d^2 X}{dT^2} = -\frac{GM}{c^2\,r^2} + \frac{2GM}{c^2\,r^3}\,X\,, \quad   \frac{d^2 Y}{dT^2} = - \frac{GM}{c^2\,r^3}\,Y\,,\quad  \frac{d^2 Z}{dT^2} = - \frac{GM}{c^2\,r^3}\,Z\,.
\end{equation}
With $(X, Y, Z) = r \,(x, y, z)$,  $T = r\,t$ and normalizations $G = c = 1$,  this approximate system is
\begin{equation}\label{A6}
 \frac{d^2 x}{dt^2} = - b\,(1-2\,x)\, \quad   \frac{d^2 y}{dt^2} = - b\,y\,,\quad  \frac{d^2 z}{dt^2} = - b\,z,
\end{equation}
where $b := GM/(c^2\,r)= M/r$.  As in the GR tidal approximation, this system has  a rest point with spatial coordinates $(1/2,0,0)$. Moreover,  linearization reveals a structure identical to the GR case: a one-dimensional stable manifold, a one-dimensional unstable manifold and a four-dimensional center manifold. Let us recall that  in the GR case when $r \gg M$ and $r \gg a$, $x_{\text{rp}} \to 1/2$ as in  Eq.~\eqref{T2}. This is in agreement with the notion that the rest point in our \emph{linear} approach comes about because of the acceleration of the fiducial observer. 

Taking the analysis one step further by expanding to higher orders in the space variables reveals that there is a rest point if the order of expansion is odd and no rest point exists when the order of expansion is even.  Note also that reductions to subcenter manifolds for the rest point of system~\eqref{A6} trivially have first integrals and the corresponding two-dimensional flows are global centers in accordance with Lyapunov's theorem. In fact, all these approximations are Hamiltonian and, in addition, have energy integrals for the entire motion. Unfortunately, the analogy with the GR tidal equations is not complete: The exact GR tidal equations are  Hamiltonian because they are the geodesic equations in the Fermi coordinate system~\cite{Chicone:2002kb}, but the constancy of the Hamiltonian during the motion does not restrict to our approximate tidal equations.

\section*{Acknowledgments}

B. M. is grateful to M. Roshan for many helpful discussions.

\end{document}